\documentclass[pdflatex,sn-basic,iicol]{sn-jnl}

\usepackage{graphicx}%
\usepackage{multirow}%
\usepackage{amsmath,amssymb,amsfonts}%
\usepackage{amsthm}%
\usepackage{mathrsfs}%
\usepackage[title]{appendix}%
\usepackage{xcolor}%
\usepackage{textcomp}%
\usepackage{manyfoot}%
\usepackage{booktabs}%
\usepackage{algorithm}%
\usepackage{algorithmicx}%
\usepackage{algpseudocode}%
\usepackage{listings}%
\usepackage{bm}
\usepackage{dsfont}

\begin{document}

\title[Article Title]{Binary AddiVortes: (Bayesian) Additive Voronoi Tessellations for Binary Classification with an application to Predicting Home Mortgage Application Outcomes}

\author[1]{\fnm{Adam J.} \sur{Stone}}\email{adam.stone2@durham.ac.uk}

\author[1]{\fnm{Emmanuel} \sur{Ogundimu}}\email{emmanuel.ogundimu@durham.ac.uk}

\author[1]{\fnm{John Paul} \sur{Gosling}}\email{john-paul.gosling@durham.ac.uk}

\affil[1]{\orgdiv{Department of Mathematical Sciences}, \orgname{Durham University}, \orgaddress{ \country{UK}}}


\abstract{The Additive Voronoi Tessellations (AddiVortes) model is a multivariate regression model that uses multiple Voronoi tessellations to partition the covariate space for an additive ensemble model. In this paper, the AddiVortes framework is extended to binary classification by incorporating a probit model with a latent variable formulation. Specifically, we utilise a data augmentation technique, where a latent variable is introduced and the binary response is determined via thresholding. In most cases, the AddiVortes model outperforms random forests, BART and other leading black-box regression models when compared using a range of metrics. A comprehensive analysis is conducted using AddiVortes to predict an individual’s likelihood of being approved for a home mortgage, based on a range of covariates. This evaluation highlights the model's effectiveness in capturing complex relationships within the data and its potential for improving decision-making in mortgage approval processes.}

\maketitle

\section{Introduction}\label{sec:intro}

Binary classification, a fundamental task in supervised learning, plays a pivotal role in extracting meaningful insights from complex datasets. Accurate binary classification models are crucial for numerous applications, including medical diagnosis \citep[for example,][]{MedicalTesting}, fraud detection \citep[for example,][]{FraudDetection}, and predicting equipment failure \citep[for example,][]{EquipmentFailure}. The increasing complexity and volume of data have driven researchers to explore innovative methods that can handle high-dimensional data, non-linear relationships and noisy environments effectively.

The AddiVortes model, introduced in \cite{AddiVortes}, is a multivariate regression algorithm that uses Voronoi tessellations and an additive approach to model a function $f$ that relates the continuous variable $Y$ to some or all $p$ potential covariates $x = (x_1, \ldots, x_p)$, such that
\begin{equation}\label{Regression}
    Y = f (x) + \epsilon, ~~~~~ \epsilon \sim N (0, \sigma^2).
\end{equation}
Voronoi tessellations and many of their applications are described in \cite{VoronoiTess}. They are a powerful mathematical tool used to partition a space into regions based on the distance to a specific set of points. By integrating this with an additive model, AddiVortes can capture complex interactions between covariates.

The AddiVortes model follows a similar approach to Bayesian additive regression trees \citep[BART,][]{BARTpaper} as it employs a Markov Chain Monte Carlo (MCMC) backfitting technique similar to the one described in \cite{MCMCBackfitting}, which allows for efficient estimation of model parameters. Additionally, regularisation priors similar to those used in \cite{CART} help reduce tessellation complexity and prevent overfitting. Many variants of the AddiVortes model for alternative cases are being developed, such as H-AddiVortes \citep{Hetero} which extends the model to the heteroscedastic setting.

In this paper, we extend the AddiVortes method with a probit model set-up for binary classification (\(Y = 0\) or \(1\)). The probit model is well-suited for binary outcomes and has been widely used in various applications due to its interpretability and performance (\cite{Albert_and_Chib}). The probit model assumes that there is an underlying continuous latent variable that follows a normal distribution, which is then thresholded to produce the binary outcome. 

The integration of AddiVortes with the probit model improves its applicability in binary classification tasks by using the strengths of both approaches. The additive approach of AddiVortes allow flexible modelling of complex relationships between covariates, while the probit model provides a robust and long-established framework for binary outcomes. This extension improves predictive performance and reliability across diverse datasets.

This advancement has significant implications for real-world applications: we will demonstrate this by predicting loan approval outcomes using data from the Home Mortgage Disclosure Act (HMDA). Enacted in 1975, HMDA was established in the America to promote transparency in mortgage lending practices. It requires financial institutions to maintain, report, and publicly disclose loan-level information about mortgages. The primary goal of HMDA is to provide the public with sufficient data to help identify potential discriminatory lending practices, ensure that financial institutions are meeting the housing needs of their communities, and facilitate the allocation of public and private investment in housing.

The data collected under HMDA is extensive, encompassing various attributes of both loan applications and the applicants themselves. This data includes information on the loan type, amount, property value, applicant demographics (such as race, gender, and income), and the outcome of the application (whether it was approved or denied). The comprehensive nature of this data allows for detailed analysis of lending patterns and practices.

The Home Mortgage Disclosure Act (HMDA) data has been extensively utilised in research to explore various aspects of mortgage lending practices. For instance, it has been used to analyse whether minority applicants face adverse lending outcomes in the mortgage market, as discussed by \citep{HMDR_Data_Minority_Applicants}; study racial and ethnic disparities in mortgage lending \citep{HMDR_Racial_and_Ethnic_Disparities} and to investigate potential discrimination by banks in lending practices \citep{HMDR_Data_lending_discrimination}.

Given the huge amount of information contained within HMDA data, it is particularly valuable for predictive analytics. Predicting whether a loan application will be accepted is not only beneficial for financial institutions but also for prospective borrowers and policymakers.

Financial institutions can use predictive models to strength decision-making processes, reduce processing times, and improve customer service. For borrowers, understanding the factors that influence loan approval can significantly improve their chances of success, enabling them to tailor their applications accordingly. On a broader scale, the insights gained from predicting loan outcomes using HMDA data can inform policy decisions aimed at promoting fair lending practices and ensuring equitable access to credit.

The remainder of the paper is organised as follows. In Section~\ref{sec:Brief Of AddiVortes}, we give a short review of the AddiVortes model. In Section~\ref{Binary_Outcomes}, we describe the probit model set-up for AddiVortes in detail, including the mathematical formulation and implementation specifics. The use of Voronoi tessellations in binary classification is highlighted in a simulated experiment in Section \ref{Simulated_study}. Section~\ref{benchmark} illustrates the potential of AddiVortes for classification through a diverse range of examples, on real-world data. We evaluate the model's performance against other established methods to highlight its strengths and weaknesses. In Section~\ref{in-depth}, we present an in-depth analysis of predicting loan application outcomes using HMDA data. Finally, Section~\ref{discussion} concludes the paper with a discussion on the findings, implications for practice, and future research directions.

\section{Review of AddiVortes}\label{sec:Brief Of AddiVortes}

\begin{figure*}[ht]
    \centering
    \includegraphics[trim = 3.5cm 4cm 2cm 3cm,clip]{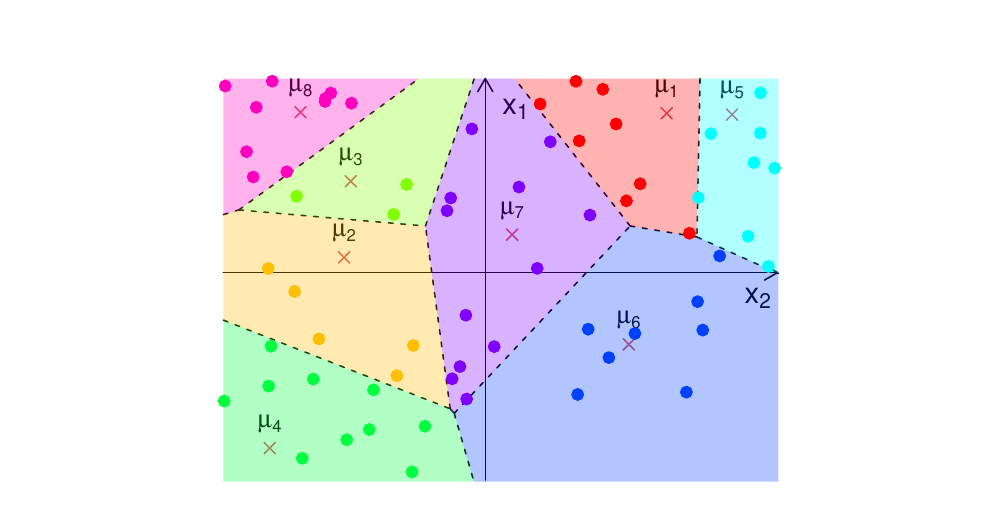}
    \caption{Example of predictive modelling using a two-dimensional Voronoi tessellation with centres at crosses, labelled with output values $(\mu_1,\ldots,\mu_8)$ associated to the given cell. Samples are represented by points with their output value corresponding to the cell they are in. \citep{AddiVortes}.}
    \label{fig:Voronoi}
\end{figure*}

The AddiVortes model consists of a predetermined fixed number, $m$, of tessellations. The $j^{\mathrm{th}}$ tessellation structure is denoted by $T_j$, that is, which covariates are dimensions in the tessellation and the coordinates of the centres of the cells. The corresponding output values for each cell are given by $M_j=\{\mu_{1j},\ldots,\mu_{bj}\}$, where $b$ is the number of cells in tessellation $T_j$. Figure~\ref{fig:Voronoi} illustrates a Voronoi tessellation with eight cells. The output value for a sample $\bm{x}$ of tessellation $T_j$,  is given by the function $g(\bm{x}|T_j,M_j)$. An estimation of $\mathds{E}(Y|\bm{x})$ is given by the sum of all the outputs of the tessellations that $\bm{x}$ corresponds to; that is, 
\begin{equation*}
    Y=\sum\limits_{j=1}^m g(\bm{x}|T_j,M_j)+\varepsilon, ~~~~ \varepsilon \sim \mathcal{N}(0,\sigma^2).
\end{equation*}
In single-dimension tessellations, the $\mu_{ij}$ represent main effects since $g(\bm{x}|T_j,m_j)$ only depends on a single covariate but will represent interaction effects when the tessellations are multi-dimensional. Thus, AddiVortes can capture both the main effects and interaction effects in the model.

The aim of AddiVortes is to extract the posterior distribution of all unknown parameters in the sum-of-tessellations model,
\begin{equation*}
    \pi((T_1, M_1), \ldots, (T_m, M_m), \sigma|\bm y_{\text{train}},\bm x_{\text{train}}).
\end{equation*}
Thus, priors are specified for the parameters of the sum-of-tessellations model, specifically  for $(T_1, M_1),\ldots,(T_m, M_m)$, and $\sigma$. These priors are strategically chosen to favour less complex configurations with fewer dimensions and cells. This regularisation controls the influence of individual tessellation effects, preventing any one of them from becoming dominant. Without these regularisation priors, complex tessellations with a large number of dimensions and centres would cause over-fitting and limit the advantages of the additive model in both approximation and computation. In this paper, we use the similar priors as the ones defined in \cite{AddiVortes}, these are explained in Section \ref{Binary_Outcomes}.

The specification of the regularization priors is simplified when independence restrictions are applied, such that,
\begin{equation}
    \begin{aligned}
        \pi(\left(T_1,\bm{M}_1\right),\ldots, & \left(T_m,\bm{M}_m\right), \sigma)\\ &=\prod\limits_{j=1}^m\left[\pi(T_j,\bm{M}_j) \right]\pi(\sigma) \\
        &=\prod\limits_{j=1}^m \left[\pi\left(\bm{M}_j|T_j\right)\pi(T_j)\right]\pi(\sigma)
    \end{aligned}
    \label{independence}
\end{equation}
and
\begin{equation}
    \pi(\bm{M}_j|T_j)=\prod\limits_{i=1}^{b_j} \pi\left(\mu_{ij}|T_j\right),
    \label{independence2}
\end{equation}
where $\mu_{ij} \in \bm{M}_j$.

A Gibbs sampler is used which involves $m$ successive draws of $(T_j, M_j)$ conditionally on $(\{T_{j'}, M_{j'}\}_{j \neq j'}, \sigma, y)$ for $j = 1, \ldots, m$, followed by a draw of $\sigma$ from its full conditional distribution. \cite{MCMCBackfitting} previously explored a similar application of the Gibbs sampler for posterior sampling in additive and generalised additive models, with $\sigma$ held fixed.

The draw of $\sigma$ from the full conditional can simply be achieved by sampling from an inverse gamma distribution and we can carry out each draw from $(T_j, M_j) ~|~(\{T_{j'} , M_{j'}\}_{j \neq j'}, \sigma, y)$ in two successive steps as
\begin{equation}
\begin{aligned}
    &T_j |\{T_{j'} , M_{j'}\}_{j \neq j'},\sigma, y\\
    &M_j |T_j ,\{T_{j'} , M_{j'}\}_{j \neq j'},\sigma, y.
\end{aligned}
\label{steps}
\end{equation}

The draw of $T_j$ in \eqref{steps} can be obtained using the Metropolis–Hastings (MH) algorithm similar to the one proposed by \cite{CART}. Six moves are suggested to propose a new tessellation based on the current tessellation, each with its associated proposal probability
\begin{itemize}
  \item adding/removing a centre (0.2 each),
  \item adding/removing a covariate (0.2 each),
  \item swapping a covariate (0.1) or
  \item changing the position of a centre (0.1).
\end{itemize}
Subsequently, the draw of $M_j$ in Equation~\eqref{steps} involves independent draws of $\mu_{ij}$ from a normal distribution since we use a conjugate normal distribution, as described in Section~\ref{Binary_Outcomes}.

By running the algorithm for a sufficient number of iterations after a suitable burn-in period, each iteration can be treated as an approximate, dependent sample from $\pi(f|y)$. To estimate $f(x)$, we can take the average or median of the posterior samples. A reasonable $(1 - \alpha)\%$ posterior interval for $f(x)$ is the interval between the upper and lower $\alpha/2$ quantiles and these uncertainty intervals behave sensibly, widening at $x$ values far from the data.

\section{Binary outcomes}\label{Binary_Outcomes}

In this section, we extend AddiVortes from the continuous response setting to binary classification $(Y=0$ or $1)$ via a probit model. The model then estimates the probability that an observation will fall into the target class $(Y=1)$. That is, the model estimates the function $p: \mathds{R}^p \mapsto [0,1]$, such that 
\begin{equation}\label{targetProb}
    p(x) \equiv P[Y = 1 | x] = \Phi[G(x)],
\end{equation}
where
\begin{equation*}
    G(x) \equiv \sum_{j=1}^{m} g(x;T_j, M_j),
\end{equation*}
and $\Phi[\cdot]$ is the standard normal cumulative distribution function. Note, the target classification probability $p(\bm x)$ is obtained as a function of the sum-of-tessellations, in contrast to aggregate classifier approaches.

To extend AddiVortes utilising Equation \ref{targetProb}, we use a regularisation prior on $G(x)$ and implement a similar Bayesian backfitting algorithm for posterior computation to one presented in \cite{AddiVortes}. We assume prior independence similar to \ref{independence} and \ref{independence2} but with $\pi(\sigma)=1$ since the model implicitly assumes $\sigma = 1$, so only a prior on $(T_1, M_1), \ldots, (T_m, M_m)$ is needed. The prior for the tessellation structure $T_j$ is specified by multiple factors:
\begin{enumerate}
  \item the number of covariates considered in $T_j$, 
  \item the number of centres in $T_j$,
  \item the covariates that are included in $T_j$ and
  \item the coordinates of the centres.
\end{enumerate}
We apply the same probability densities as in \cite{AddiVortes}.

For $p(\mu_{ij} | T_j)$, a conjugate normal prior (conjugate because $\epsilon$ in \ref{Regression} follows a normal distribution) is considered
\begin{equation}\label{mu prior}
    \mu_{ij} \sim \mathcal{N}(0, \sigma_{\mu}^2),
\end{equation}
where $\sigma_{\mu} = \frac{3.0}{k \sqrt{m}}$, and $k$ is chosen such that $G(x)$ will likely be in the interval $(-3.0, 3.0)$. The prior shrinks tessellation output values $\mu_{ij}$ toward zero, limiting the effect of individual tessellation of $G(x)$.

By shrinking $G(x)$ toward 0, the prior \ref{mu prior} has the effect of shrinking $p(x) = \Phi[G(x)]$ toward 0.5. If it is of interest to shrink to a value $p_0$ other than 0.5, one can simply replace $G(x)$ by $G_c = G(x)+ c$ in \ref{targetProb} with the offset $c = \Phi^{-1}[p_0]$. Note also that if an interval other than $(\Phi[-3.0], \Phi[3.0])$ is of interest for $p(\bm x)$, suitable modification of \ref{mu prior} is straightforward.

For the posterior calculation, the backfitting algorithm in \cite{AddiVortes} can be implemented using data augmentation as described in \cite{Albert_and_Chib}. The model is recast by introducing latent variables $\bm Z = \{Z_1, \ldots,Z_n\}$, that determine the value of $Y$ as follows
\begin{equation*}
    \left\{\begin{array}{cc}
            Y_i = 1 ~~~~~ \text{if } Z_i > 0 \\
            Y_i = 0 ~~~~~ \text{if } Z_i \leq 0
        \end{array}
    \right.
\end{equation*}
Following \cite{AddiVortes} again, we assume that $\bm Z$ is related to $\bm x$ through an additive model, where each additive component is a tessellation based on the covariates. Thus, for a given observation $\bm x_i$, $Z_i$ is formally modelled as
\begin{equation*}
\begin{aligned}
    Z_i = g(x; T_1, M_1) +& g(x; T_2, M_2) + ...\\& + g(x; T_m, M_m) + \epsilon,
\end{aligned}
\end{equation*}
and we further assume that $\epsilon \sim N(0, 1)$, using a probit link.

We use the data augmentation method, similar to one used in \cite{dataMissing}, by treating $\bm Z$ as missing data, and then use the Gibbs sampler to generate samples from the posterior distribution $p\{(T_1, M_1), (T_2, M_2), . . . , (T_m, M_m), \bm Z|\bm x,\bm y\}$. The Gibbs sampler composes of drawing $n$ successive draws of $z_i$ (for $i = \{1,..., n\}$) from $p(\bm Z|(T_1, M_1), (T_2, M_2), . . . , (T_m, M_m), \bm x,\bm y)$ followed by $m$ successive draws of $(T_j, M_j)$ for $j = 1,..., m$ from $p((T_j, M_j)|\{T_{j'}, M_{j'}\}_{j \neq j'}, \bm Z, \bm x,\bm y)$ using the AddiVortes method presented in \cite{AddiVortes}. Let $\hat{z_i} =  \sum\limits_{j=1}^mg(x_i; T_j, M_j)$ denote the fitted value for observation $i$ from the $m$ trees. Then, $ z_i$ (for $i = \{1,..., n\}$) can be independently generated from truncated normal distributions
\begin{equation*}
    \left\{\begin{array}{cc}
    z_i \sim \text{max}(N(\hat{z_i},1),0) ~~ \text{if } y_i=1 \\
    z_i \sim \text{min}(N(\hat{z_i},1),0) ~~ \text{if } y_i=0 
    \end{array}\right.
\end{equation*}

After $\sigma_\mu^2$ has been chosen according to the procedure described above, and after a suitable burn-in period used to reach convergence, we can use the subsequent $K$ posterior draws for inference. Denote these $K$ posterior draws as $\{(T^{(k)}_1 , M^{(k)}_1 ), . . . , (T^{(k)}_m , M^{(k)}_m )\}^K_{k=1}$. Given the $k$th draw, the probability that an observation with input variables $x$ belongs to class 1 is $\Phi\left\{\sum\limits^m_{j=1} g(\bm x, T^{(k)}_j , M^{(k)}_j )\right\}$, where $\Phi$ is the cumulative distribution function of standard normal distribution. Therefore, the posterior average probability that an observation with input variables $x$ belongs to class 1 can be estimated as
\begin{equation}\label{Posterior Samples}
   \frac{1}{K} \sum\limits_{k=1}^{K}\Phi\left\{ \sum\limits_{j=1}^m g(x;T_j^{(k)},M_j^{(k)})\right\}
\end{equation}
We use \eqref{Posterior Samples} to classify observations, if the probability calculated from \eqref{Posterior Samples} is larger than 0.5 (or a chosen threshold value), then the observation is classified as 1; otherwise 0.

To implement Binary AddiVortes efficiently, we use the FNN (Fast Nearest Neighbor Search Algorithms and Applications) CRAN package \citep{FNN}, which utilises C++ for quick nearest-centre search. The regularisation priors also means tessellations have lower dimensions and fewer centres, reducing computational complexity in nearest-centre search. As a result, Binary AddiVortes achieves competitive runtimes compared to BART and other black-box binary classification methods.

\section{Simulation study}\label{Simulated_study}

In this section, the strengths of AddiVortes in comparison to tree-based models such as BART are illustrated through two simulated experiments using synthetic functions: the \textit{rotated axis function} and the \textit{sinusoid function} from \cite{oblique}. These functions were selected to evaluate the model's ability to handle both oblique decision boundaries and highly non-linear patterns, which are common in classification tasks.

\begin{figure*}[ht]
    \centering
    \includegraphics[width=15cm]{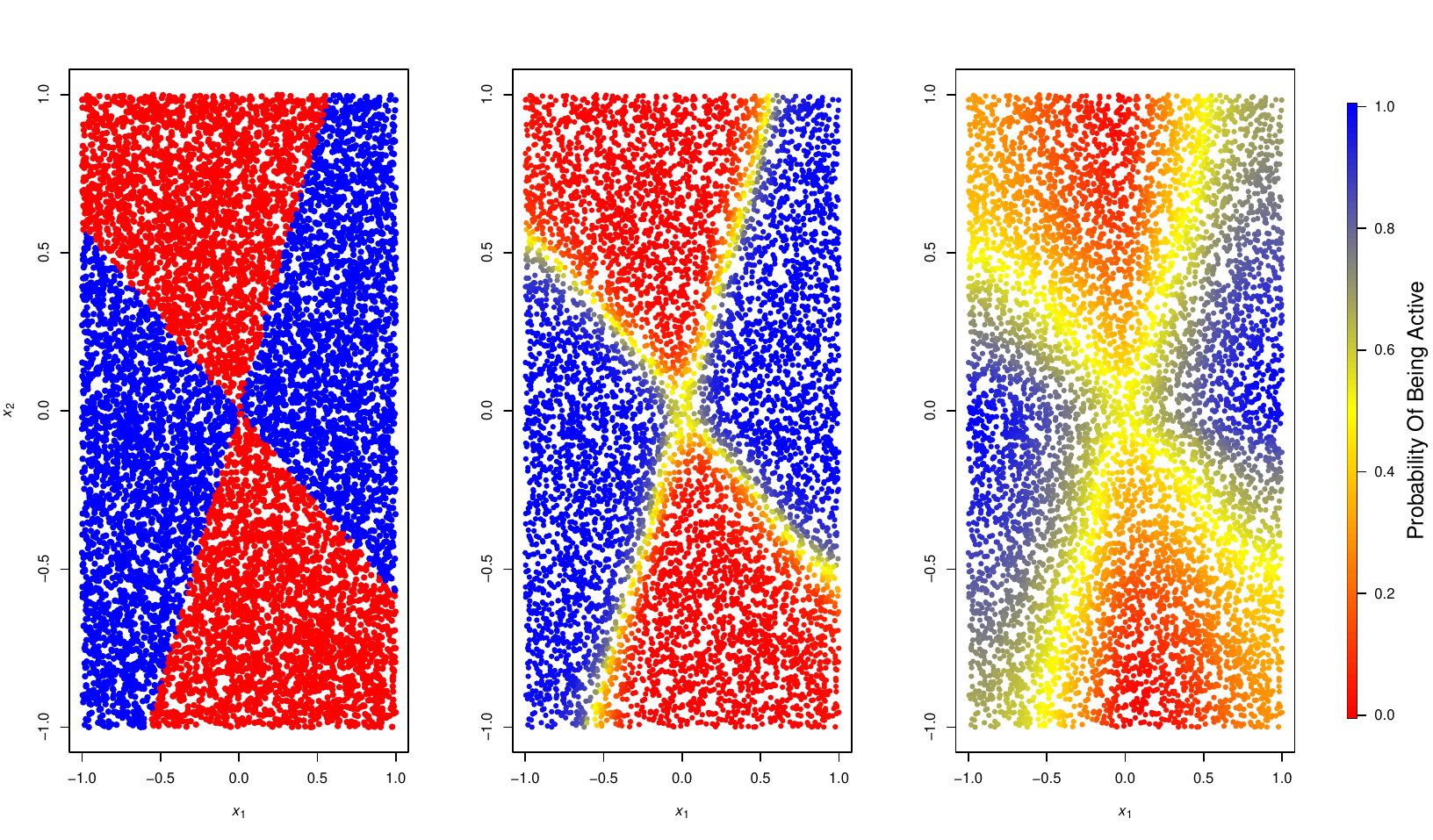}
    \caption{The rotational axis function with $\theta = \pi/6$ (left); the predicted probability of rotated axis function for AddiVortes (middle) and BART (right). }
    \label{fig:rotated_function}
\end{figure*}

The \textbf{rotated axis function} is designed to test a model’s capability to approximate decision boundaries that are not aligned with the coordinate axes. For a given point \( \bm{x} = (x_1, x_2) \), the function first applies a counter-clockwise rotation by an angle \( \theta \) and then classifies the point based on its location relative to the rotated axes. Mathematically, the rotation is defined as
\[
u(\bm{x}) = 
\begin{bmatrix}
u_1 \\ u_2
\end{bmatrix}
=
\begin{bmatrix}
\cos(\theta) & -\sin(\theta) \\
\sin(\theta) & \cos(\theta)
\end{bmatrix}
\begin{bmatrix}
x_1 \\ x_2
\end{bmatrix},
\]
where \( u_1 \) and \( u_2 \) are the rotated coordinates of \( \bm{x} \). The rotated axis function \( f(\bm{x}; \theta) \) is then defined as
\[
f(\bm{x}; \theta) = 
\begin{cases} 
1 & \text{if } u_1 \cdot u_2 > 0, \\
0 & \text{otherwise,}
\end{cases}
\]
where \( \theta \in [0, \pi/4] \) controls the angle of rotation. This function is depicted in Figure \ref{fig:rotated_function} (left) for $\theta = \pi/6$.

\begin{figure*}[ht]
    \centering
    \includegraphics[width=15cm]{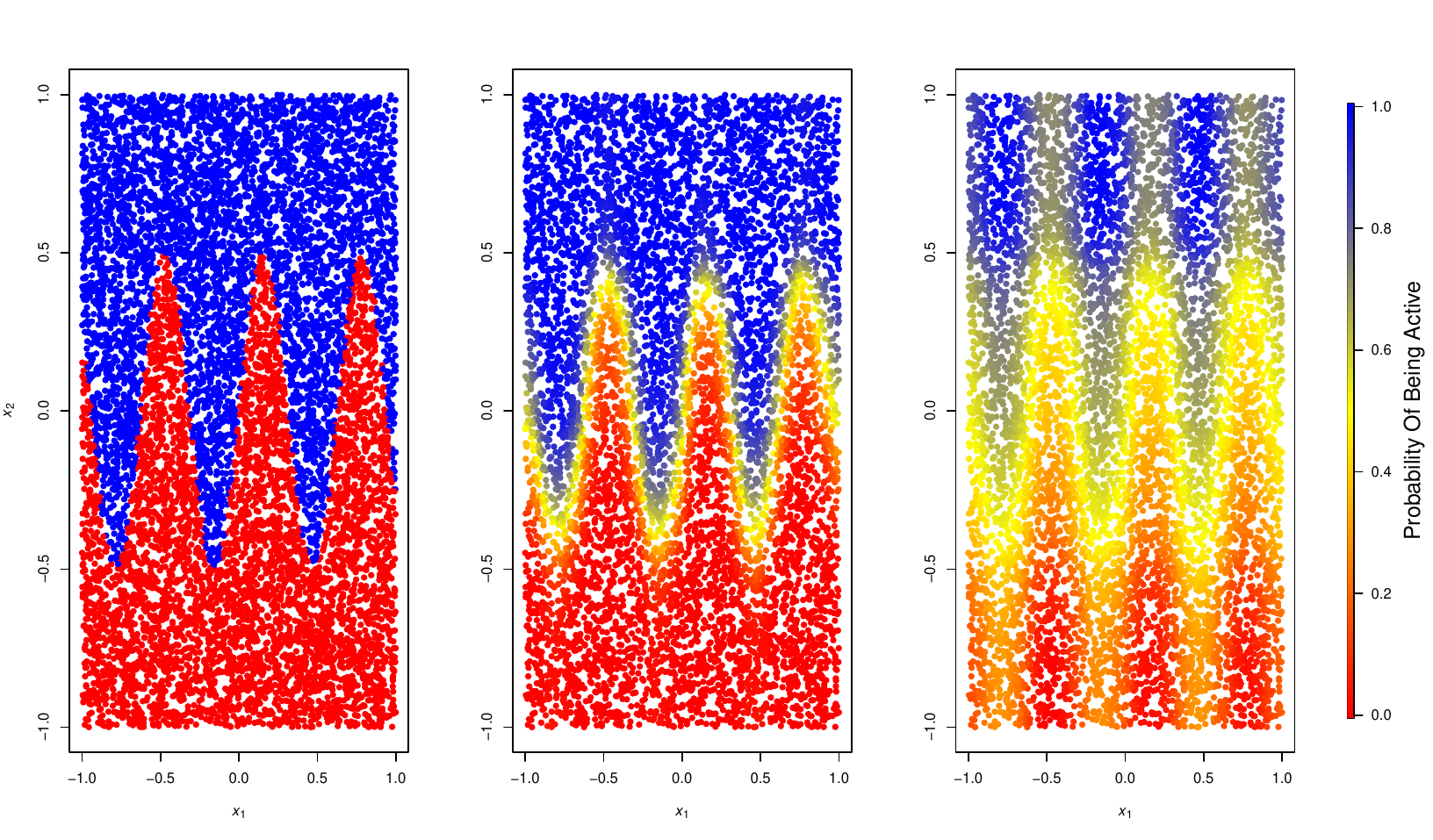}
    \caption{The sinusoid function with $\alpha=0.5$ (left); the predicted probability of sinusoid for AddiVortes (middle) and BART (right).}
    \label{fig:Sin_function}
\end{figure*}

The \textbf{sinusoid function} evaluates the model’s ability to capture complex, non-linear decision boundaries. The classification boundary is defined by a sinusoidal curve, where the outcome depends on whether the point lies above or below the curve. The sinusoid function \( f(\bm{x}; \alpha) \) is given by
\[
f(\bm{x}; \alpha) = 
\begin{cases} 
1 & \text{if } x_2 > \alpha \sin(10x_1), \\
0 & \text{otherwise,}
\end{cases}
\]
where \( \alpha \) represents the amplitude of the sinusoidal curve. This function is shown in Figure~\ref{fig:Sin_function} (left) for $\alpha =0.5$.

To demonstrate the ability of the models to approximate these functions, we sampled \( n \) two-dimensional observations \( \bm{x}_i \sim \text{Uniform}([0, 1]^2) \), and generated the dependent variables \( y_i \) based on the rotated axis function \( f(\bm{x}; \theta) \) or the sinusoid function \( f(\bm{x}; \alpha) \). For both functions, we applied the AddiVortes model and the BART model to perform binary classification. 

Figures~\ref{fig:rotated_function} (middle) and \ref{fig:Sin_function} (middle) illustrate the predicted probabilities for \( Y = 1 \) using AddiVortes, while Figures~\ref{fig:rotated_function} (right) and \ref{fig:Sin_function} (right) present the results for BART. As shown, AddiVortes demonstrates greater certainty near the boundaries of the decision functions, whereas BART struggles to maintain consistency in these regions.

\begin{figure*}[ht]
    \centering
    \includegraphics{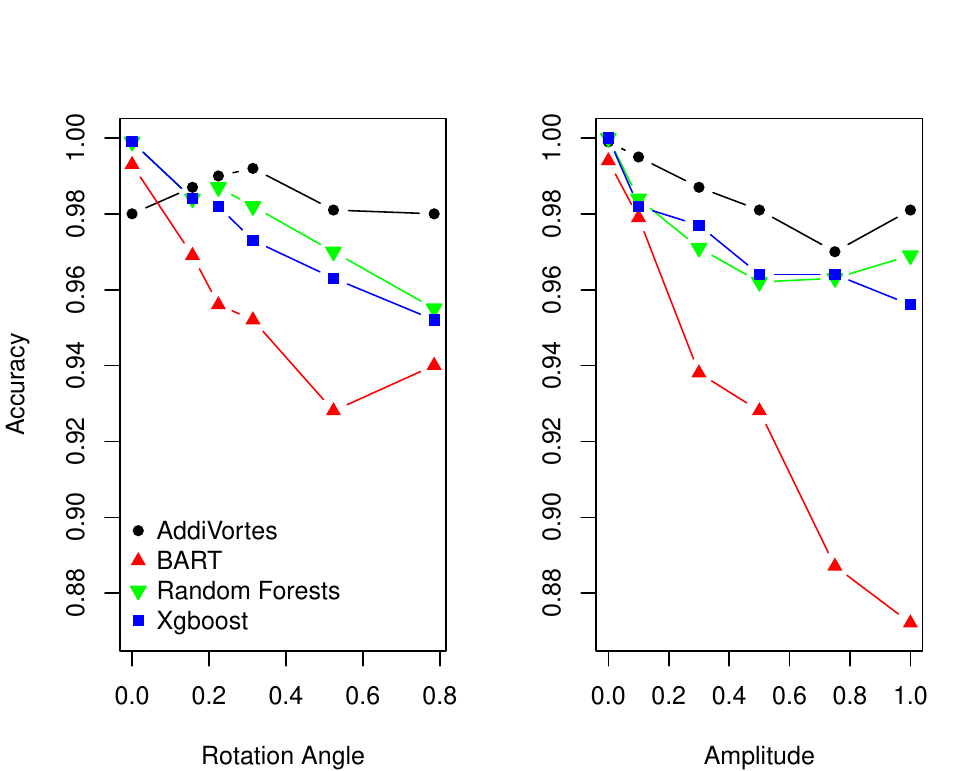}
    \caption{The accuracy of prediction for AddiVortes and competing methods for the rotated axis function for varying $\theta$ (left) and the sinusoid function for varying $\alpha$.}
    \label{fig:Accuracy_rot_sin}
\end{figure*}

To further illustrate the performance of the models, Figure \ref{fig:Accuracy_rot_sin} summarises the accuracy of both AddiVortes and BART for varying values of the parameters \( \theta \) (for the rotated axis function) and \( \alpha \) (for the sinusoid function). 1,000 observations where generated for each of the training and testing sets. The left panel of Figure \ref{fig:Accuracy_rot_sin} shows the accuracy as a function of the rotation angle \( \theta \), while the right panel displays the accuracy as a function of the amplitude \( \alpha \). 

As competitors, we evaluated four black-box methods: random forests \citep[][implemented as \texttt{randomforest} in R]{RandomForests}, (extreme) gradient boosting \citep[][implemented as \texttt{XGboost} in R]{XGboost} and BART \citep[][implemented as \texttt{bart} from the BayesTree R package]{BARTpaper}. These models were selected for their robust binary classification capabilities, interpretability and comparability.

In the case of the rotated axis function, AddiVortes consistently achieves higher accuracy across all values of \( \theta \), demonstrating its robustness in capturing oblique decision boundaries. Notably, the performance of BART decreases as \( \theta \) increases, highlighting its difficulty in modelling decision boundaries that deviate significantly from the coordinate axes. This result underscores the advantage of the additive Voronoi tessellations in handling rotated feature spaces.

For the sinusoid function, AddiVortes again outperforms BART across the range of \( \alpha \) values. The accuracy of AddiVortes remains stable even as the non-linearity of the boundary increases with larger values of \( \alpha \). In contrast, BART shows noticeable performance degradation for higher amplitudes, indicating its limited ability to adapt to highly non-linear decision boundaries.

\section{Benchmark Data Sets}\label{benchmark}

We compared the predictive performance of AddiVortes against several competing algorithms for six benchmark datasets, five from the UCI Machine Learning Repository: Breast Cancer, Ionosphere, Sonar, Heart Failure and German credit and one from the IEEEdataPort on digital sensors detecting oil or water. These data sets exhibit varying sample sizes, ranging from 208 to 4,475 observations, with each data set having a binary output variable and between 9 and 60 covariates. 

As metrics, we use the accuracy as defined in Section \ref{Simulated_study} and the Area Under the Curve (AUC) values. The AUC is a metric used to evaluate the performance of binary classification models by measuring the area under the Receiver Operating Characteristic (ROC) curve. The ROC curve plots the true positive rate against the false positive rate across different classification thresholds. An AUC value of 0.5 indicates random guessing, while a value of 1 represents perfect classification.

\begin{table}[ht]
  \centering
  \begin{tabular}{llll}
    \hline
    \textbf{Name} & \bm{$n$} & \textbf{TC(\%)} & \textbf{\# Cov} \\
    \hline
    Sonar  & 208 & ~~~~~~~~53.4\%&  ~~~~~~~60 \\
    Heart Failure  & 299 & ~~~~~~~~32.1\%&  ~~~~~~~12 \\
    German Credit  & 1000 & ~~~~~~~~30.0\%&  ~~~~~~~24 \\
    Breast Cancer  & 683 & ~~~~~~~~35.0\%&  ~~~~~~~9 \\
    Digital Sensors  & 4475 & ~~~~~~~~57.4\%&  ~~~~~~~10 \\
    Ionosphere & 351 & ~~~~~~~~64.1\%&  ~~~~~~~32 \\
    \hline    
  \end{tabular}
  \caption{A table showing the data sets used in our analysis, the percentage of observations in the target class and number of covariates in the datasets.}
  \label{tab:datasets}
\end{table}

\begin{table*}[ht]
  \centering
  \begin{tabular}{lll}
    \hline
    \textbf{Method} & \textbf{Parameters} & \textbf{Values considered} \\
    \hline
    Random forests &  Number of trees & 500 \\
    &\% variables sampled to grow each node & 10, 25, 50, 100 \\
    \hline
    XGBoost &  Number of trees & 50, 100, 200 \\
        &Shrinkage (multiplier of each tree added)& 0.01, 0.05, 0.10, 0.25\\
        &Max depth permitted for each tree&  3, 5, 7 \\
        \hline
    BART 
        &Number trees, m &50, 200 \\
        &$\mu$ prior: $k$ value for $\sigma_\mu$ &1, 2, 3, 4, 5\\
        \hline
    AddiVortes & \# Tessellations: m & 200 \\
        &$\mu$ prior: $k$ value for $\sigma_\mu$ & 3\\
        & Standard deviation of centre location: $\sigma_c$ &  0.2, 0.4 \\
        & Probability weight for \# covariates: $\omega$ & 3, 5\\
        & Poisson rate for \# centres: $\lambda_c$ & 15, 30, 45\\
        
    \hline
  \end{tabular}
  \caption{A table showing the cross-validation values for competing methods.}
  \label{tab:CVvalues}
\end{table*}

For each dataset, we generated 20 independent train/test split samples (sampling without replacement), allocating 80\% of the data to the training set and 20\% to the test set. All models were subjected to five-fold cross-validation to optimise hyperparameters, with the ranges of values considered for each method detailed in Table~\ref{tab:CVvalues}.

Tables \ref{tab:Accuracy} and \ref{tab:AUC} present the accuracy and AUC values of AddiVortes compared to competing methods. These results highlight the exceptional performance and robustness of AddiVortes in binary classification tasks. 

\begin{table}[ht]
  \centering
  \begin{tabular}{lllll}
    \hline
    \textbf{Dataset} & AV  &BART& RF   & XGBoost\\
    \hline
    Sonar  & \textbf{86.4}  &  75.7 & 81.3  & 83.8 \\
    Heart Failure  & \textbf{85.8}  &  80.3 & 85.3  & 83.8\\
    Breast Cancer  & \textbf{97.2} & 96.6 &  \textbf{97.2}  & 96.3\\
    German Credit  & 76.4 &  76.1 & \textbf{76.5}  & 75.9 \\
    Ionosphere  & \textbf{93.7}  &  88.7 & 93.2  & 91.8 \\
    Digital Sensor  & \textbf{99.7} & 96.0 & 99.0  &99.7 \\
    \hline    
  \end{tabular}
  \caption{A table showing the Accuracy(\%) of each method for the datasets.}
  \label{tab:Accuracy}
\end{table}

From Table \ref{tab:Accuracy} shows that AddiVortes achieves the highest accuracy in four out of six datasets, demonstrating its adaptability to diverse scenarios.  The results demonstrate its adaptability to diverse data structures and its competitiveness against established methods such as BART, XGBoost, and RF. AddiVortes performs particularly well in structured datasets and remains highly competitive even when not achieving the top accuracy. Its ability to maintain strong performance across varied datasets underscores its robustness and versatility in binary classification tasks.

\begin{table}[ht]
  \centering
  \begin{tabular}{lllll}
    \hline
    \textbf{Dataset} & AV  &BART& RF   & XGBoost\\
    \hline
    Sonar  & \textbf{95.2}  &  92.1 & 94.4  & 93.4 \\
    Heart Failure  & \textbf{91.8}   &  91.5 & 91.6  & 89.7\\
    Breast Cancer  & \textbf{99.4} & 99.3 &  99.1 & 99.2\\
    German Credit  & \textbf{79.3}  &  78.6 & 78.9  & 78.3 \\
    Ionosphere  & \textbf{98.0}  &  97.6 & 97.4  & 96.7 \\
    Digital Sensor  & \textbf{100.0} & 99.8 & 99.9  &99.9 \\
    \hline    
  \end{tabular}
  \caption{A table showing the AUC of each competitor on the datasets.}
  \label{tab:AUC}
\end{table}

Table \ref{tab:AUC} provides additional insights into the discriminative ability of the models through their AUC values. AddiVortes achieves the highest AUC on all six of the datasets which is not too surprising as AddiVortes is able to partition the covariate space in a more flexible way then the competitors that use decision trees. These results highlight its superior ability to rank positive and negative classes correctly across various thresholds. These results further confirm the reliability and adaptability of AddiVortes in datasets with varying complexities and imbalances.

The results from Tables \ref{tab:Accuracy} and \ref{tab:AUC} demonstrate the strength of AddiVortes in delivering high accuracy and robust classification performance across a diverse range of datasets. Its consistent top-tier performance establishes it as a reliable and generalisable tool for predictive modelling, particularly in scenarios demanding interpretability and the ability to model complex data structures effectively.

While other methods exhibited variability in their rankings across datasets, AddiVortes consistently delivered competitive results, making it a reliable choice for binary classification tasks. These findings establish AddiVortes as a powerful and generalisable tool for predictive modelling, particularly in scenarios that demand interpretability and the ability to model complex data structures.

\section{Home Mortgage Acceptance Predictions}\label{in-depth}

We use the ``One Year National Loan Level Dataset from 2022 and 2023'' published under the Home Mortgage Disclosure Act by the FFIEC (Federal Financial Institution Examination Council), which contains Mortgage information of over a million applicants of home mortgages in America. In this paper, we train the models using the 2022 dataset to predict if a single person applicant will be accepted or rejected for a mortgage based on a number of features listed in Table \ref{FeaturesTable}, and test the models using the 2023 mortgage data. We have preprocessed the data by removing any applicants that having missing data and all features that are only relevant to joint applicants. 

To reduce computational time, we sampled
20,000 applicants to be used in for training and testing sets. To compare methods on the home mortgage dataset, we use the same competitors as in Section~\ref{benchmark}, tuning the hyperparameters by sampling five distinct sets of 20,000 samples from the 2022 mortgage data to train the models and 20,000 samples from the 2023 data, and choosing the hyperparameters from Table~\ref{tab:CVvalues} that give the best accuracy values.

\begin{table*}[ht]
  \centering
  \fontsize{8}{8}\selectfont
  \begin{tabular}{p{4.4cm}p{10cm}}
    \toprule
    \textbf{Variable} & \textbf{Description} \\
    \midrule
    
    derived\_loan\_product\_type & Derived loan product type from Loan Type and Lien Status fields for easier querying of specific records \\
    derived\_dwelling\_category & Derived dwelling type from Construction Method and Total Units fields for easier querying of specific records \\
    derived\_race & Single aggregated race categorisation derived from applicant \\
    derived\_sex & Single aggregated sex categorisation derived from applicant \\
    purchaser\_type & Type of entity purchasing a covered loan from the institution \\
    conforming\_loan\_limit & Indicates whether the reported loan amount exceeds the GSE (government sponsored enterprise) conforming loan limit \\
    preapproval & Whether the covered loan or application involved a request for a preapproval of a home purchase loan under a preapproval program \\
    loan\_type & The type of covered loan or application \\
    loan\_purpose & The purpose of covered loan or application \\
    lien\_status & Lien status of the property \\
    open-end\_line\_of\_credit & Whether the covered loan or application is for an open-end line of credit \\
    business\_or\_commercial\_purpose & Whether the covered loan or application is primarily for a business or commercial purpose \\
    loan\_amount & The amount of the covered loan, or the amount applied for \\
    combined\_loan\_to\_value\_ratio & The ratio of the total amount of debt secured by the property to the value of the property relied on in making the credit decision \\
    hoepa\_status & Whether the covered loan is a high-cost mortgage \\
    interest\_only\_payment & Whether the contractual terms include, or would have included, interest-only payments \\
    balloon\_payment & Whether the contractual terms include, or would have included, a balloon payment \\
    other\_nonamortizing\_features & Whether the contractual terms include, or would have included, any non-standard term that would allow for payments other than fully amortizing payments during the loan term \\
    property\_value & The value of the property securing the covered loan relied on in making the credit decision \\
    occupancy\_type & Occupancy type for the dwelling \\
    total\_units & The number of individual dwelling units related to the property \\
    income & The gross annual income, in thousands of dollars \\
    debt\_to\_income\_ratio & The ratio, as a percentage, of the applicant’s total monthly debt to the total monthly income \\
    applicant\_credit\_score\_type & The name and version of the credit scoring model used to generate the credit score \\
    applicant\_ethnicity & Ethnicity of the applicant \\
    applicant\_race & Race of the applicant \\
    applicant\_race\_observed & Whether the race of the applicant was collected on the basis of visual observation or surname \\
    applicant\_sex\_observed & Whether the sex of the applicant was collected on the basis of visual observation or surname \\
    submission\_of\_application & Whether the applicant submitted the application directly to the financial institution \\
    initially\_payable\_to\_institution & Whether the obligation was initially payable to the financial institution \\
    aus & The automated underwriting system (AUS) used by the financial institution to evaluate the application \\
    tract\_population & Total population in tract \\
    tract\_minority\_population\_percent & Percentage of minority population to total population for tract, rounded to two decimal places \\
    ffiec\_msa\_md\_median\_family\_income & FFIEC Median family income in dollars for the MSA/MD in which the tract is located \\
    tract\_to\_msa\_income\_percentage & Percentage of tract median family income compared to MSA/MD median family income \\
    tract\_owner\_occupied\_units & Number of dwellings, including individual condominiums, that are lived in by the owner \\
    tract\_one\_to\_four\_family\_homes & Dwellings that are built to houses with fewer than 5 families \\
    tract\_median\_age\_of\_housing\_units & Tract median age of homes \\
    \bottomrule
  \end{tabular}
  \caption{Table of the features of the home mortgage dataset.}
  \label{FeaturesTable}
\end{table*}

\begin{figure*}[ht]
    \centering
    \includegraphics{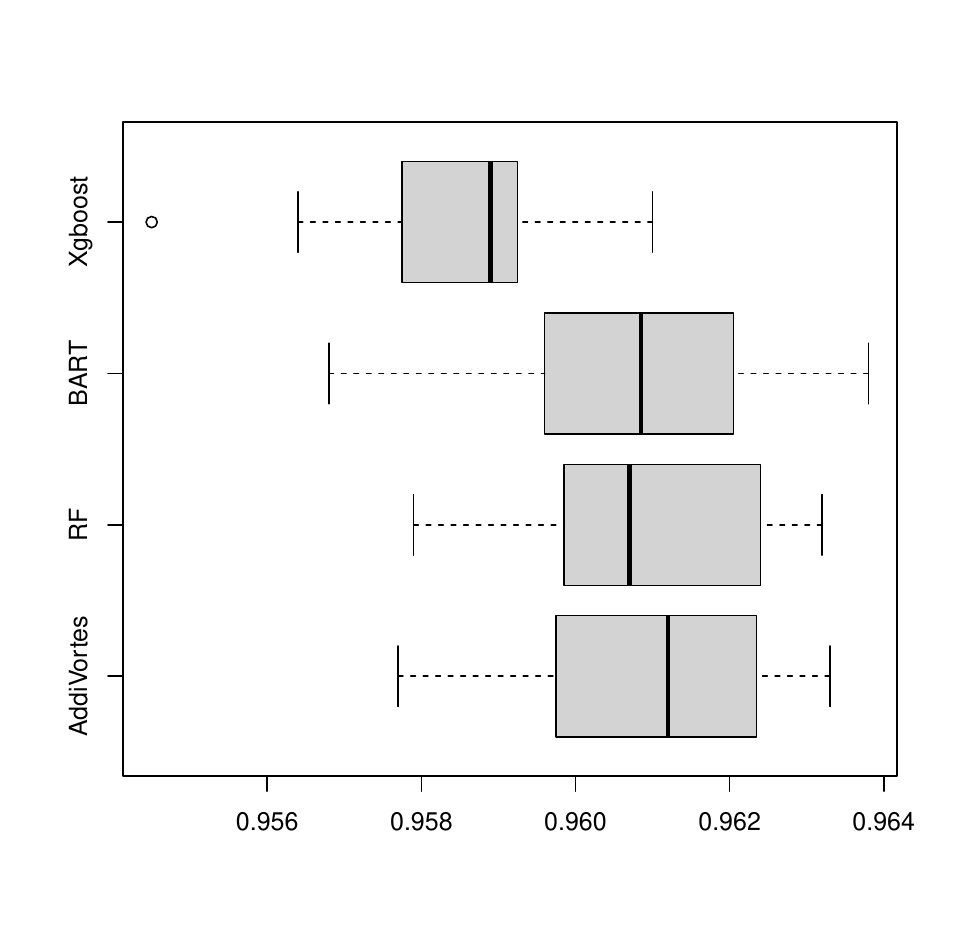}
    \caption{A boxplot of the accuracy values for each competing method on the home mortgage dataset.}
    \label{fig:Accuracy_Mortgage}
\end{figure*}

\begin{figure*}[ht]
    \centering
    \includegraphics{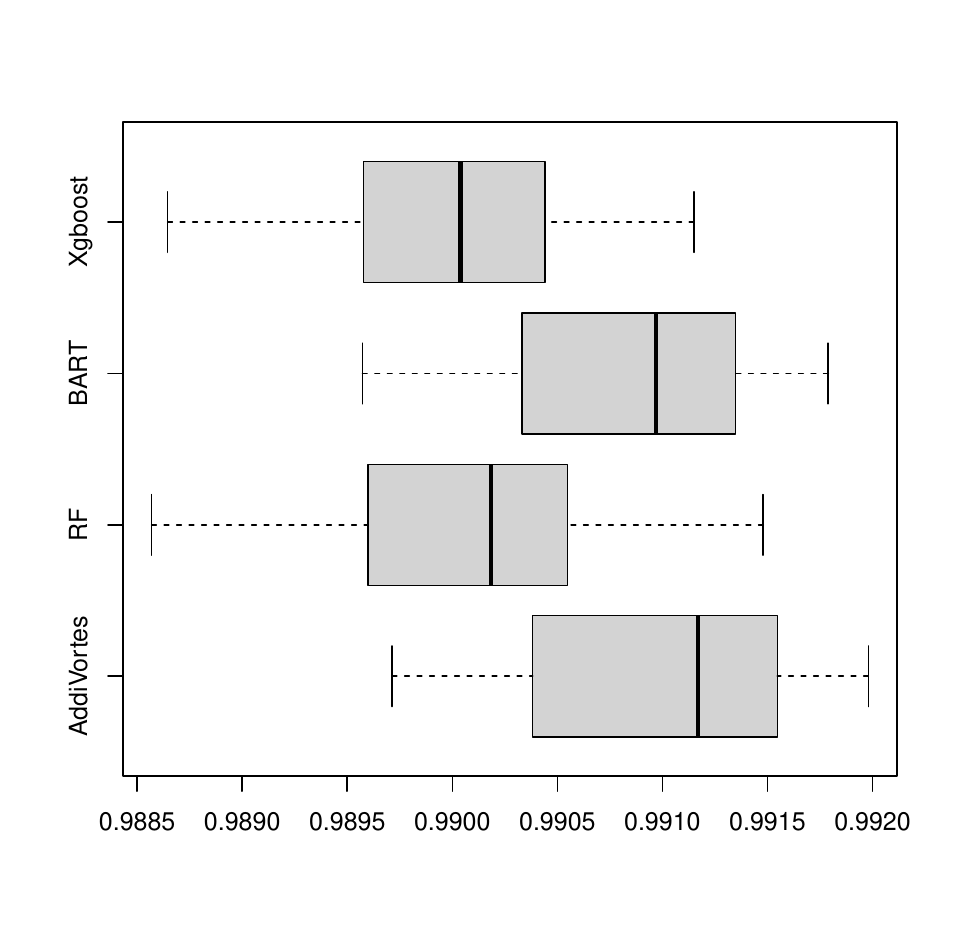}
    \caption{A boxplot of the AUC values for each competing method on the home mortgage dataset.}
    \label{fig:AUC_Mortgage}
\end{figure*}

To produce the boxplots in Figures \ref{fig:Accuracy_Mortgage} and \ref{fig:AUC_Mortgage}, we sampled 24 sets of 20,000 samples from the 2022 mortgage data to train the models and 20,000 samples from the 2023 data to obtain 24 accuracy and AUC values for each method. 

As shown in Figure \ref{fig:Accuracy_Mortgage}, AddiVortes consistently achieves the highest accuracy among all competing methods on the Home Mortgage dataset. The boxplot illustrates the variation in accuracy across multiple runs, highlighting the robustness of AddiVortes in mortgage approval prediction. The median accuracy of AddiVortes surpasses that of Random Forest (RF), Bayesian Additive Regression Trees (BART), and XGBoost, demonstrating its effectiveness in capturing complex relationships in the data. Note that the accuracy is much higher than the median base accuracy of 79\% for all methods.

Similarly, Figure \ref{fig:AUC_Mortgage} presents the area under the curve (AUC) values for the same dataset. AddiVortes achieves the highest median AUC, indicating superior discriminative power in classifying approved and rejected loan applications. The narrow spread of the AUC values for AddiVortes suggests that it maintains consistent performance across different training and testing samples. This reliability is crucial in financial decision-making, where model stability is essential.

These results further reinforce the advantages of AddiVortes in binary classification tasks, particularly in mortgage lending applications. Its ability to outperform traditional black-box models while maintaining interpretability makes it a valuable tool for financial institutions seeking to enhance their predictive analytics frameworks.

\begin{figure*}[ht]
    \centering
    \includegraphics[width=15cm]{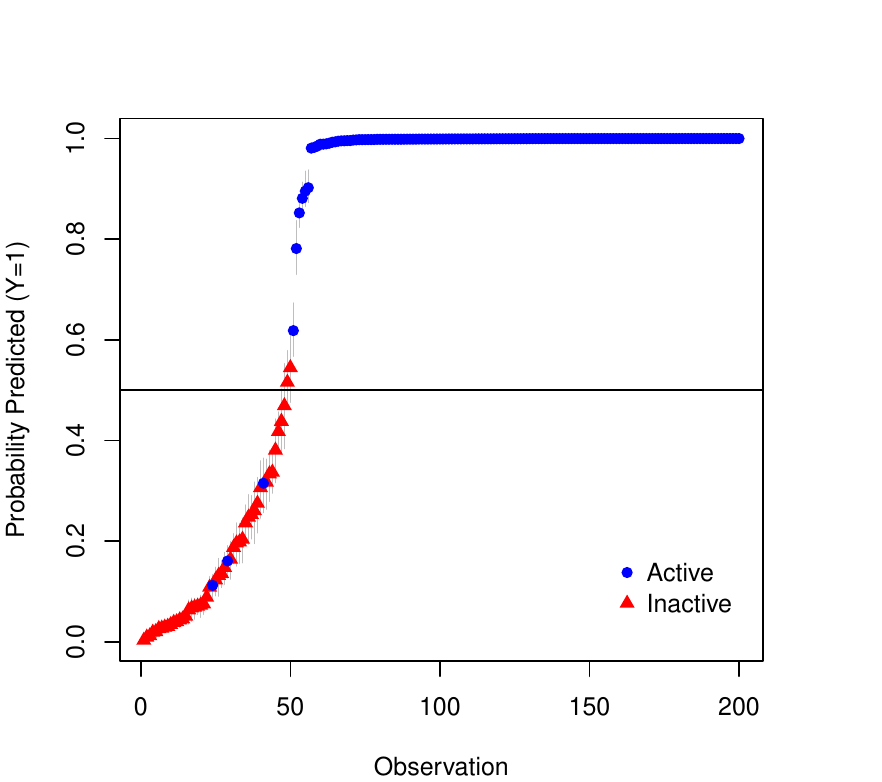}
    \caption{A graph of the predicted probabilities of an observation being active with the 90\% confidence intervals for the test set and the true status of activity.}
    \label{fig:Predicted_Probability}
\end{figure*}

Figure~\ref{fig:Predicted_Probability} illustrates the predicted probabilities \( P[Y = 1|x] \) for a test set of size 200, derived from a model trained on 20,000 observations. The graph includes 90\% posterior intervals, which convey the uncertainty in the predictions, along with markers indicating whether each loan was accepted (\( Y = 1 \)) or rejected (\( Y = 0 \)). The posterior intervals widen for observations far from the majority of the training data, reflecting increased uncertainty. Notably, the majority of misclassified observations have posterior intervals that include \( P[Y = 1|x] = 0.5 \), indicating ambiguity in these predictions. The overall test set accuracy was 97.5\%, significantly exceeding the 70\% base rate, highlighting the effectiveness of AddiVortes in predicting loan approval outcomes.

\begin{figure*}[ht]
    \centering
    \includegraphics{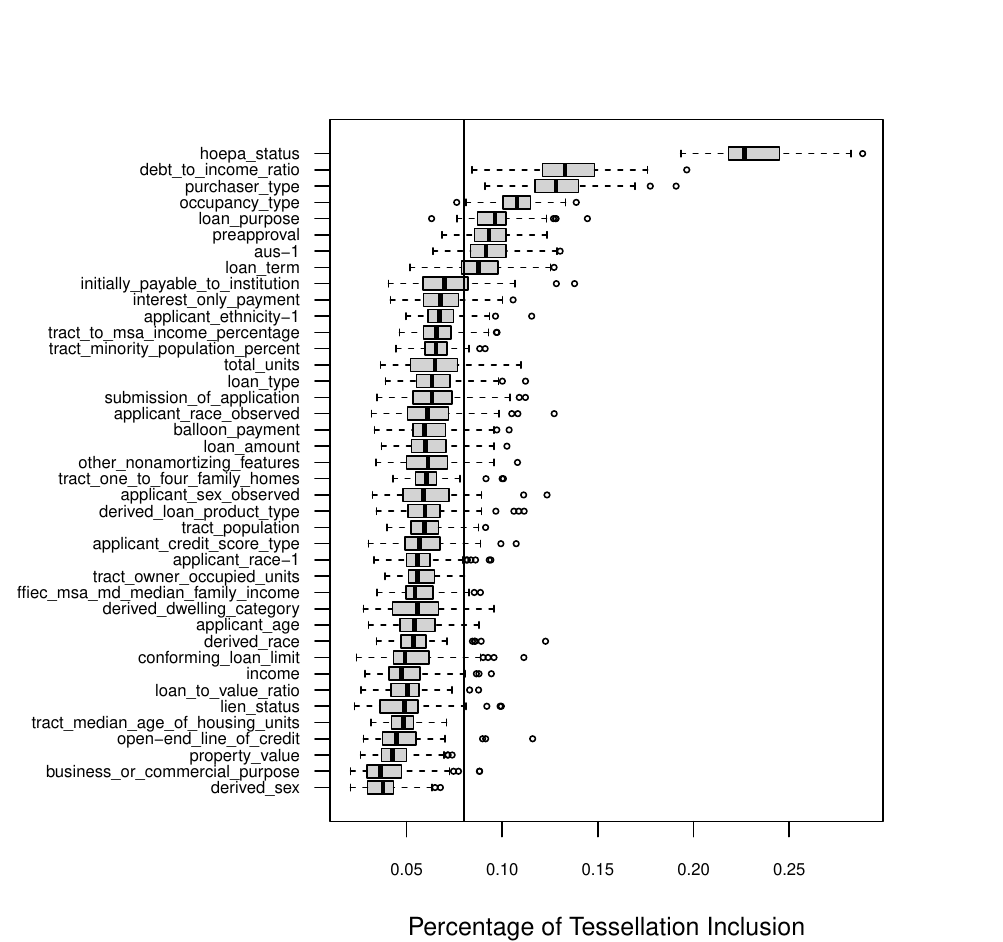}
    \caption{A boxplot showing the percentage of times a variable is included in a tessellation. A vertical line is drawn at 8\% inclusion percentage.}
    \label{fig:Variable_selection}
\end{figure*}

Figure \ref{fig:Variable_selection} provides further insight into how AddiVortes constructs its predictive model by illustrating the percentage of times each variable is included in a tessellation across 100 runs. Variables with higher inclusion rates are considered more influential in determining mortgage approval outcomes. Key predictors such as debt-to-income ratio, credit score type, and loan-to-value ratio are among the most frequently included, aligning with established financial risk assessment criteria. Additionally, tract population, minority population percentage, and applicant demographics appear frequently, suggesting that AddiVortes captures both individual financial attributes and broader socioeconomic factors.

\section{Discussion}\label{discussion}

In this paper, we extend the AddiVortes model to binary classification using a probit framework. A latent variable \( Z = f(\bm X) + \epsilon \) is introduced, where \( Y \) is determined by thresholding \( Z \). The probability estimate \( P( Y = 1 \mid \bm X) = \Phi(f(\bm X)) \) is obtained, with MCMC sampling enabling Bayesian uncertainty quantification.

Through an empirical study using mortgage approval data, we demonstrated that AddiVortes effectively models relationships between financial, demographic, and geographic covariates while capturing local structure in covariate space. The method’s ability to integrate tessellation-based feature interactions within a structured probabilistic framework offers advantages in both predictive performance and interpretability. Compared to traditional black-box models, AddiVortes provides a transparent approach that enables clear identification of influential features, which is essential for fairness in decision-making. These findings highlight its practical relevance in mortgage lending and other classification tasks where both accuracy and interpretability are critical.

Future research could extend this work to the multinomial probit regression framework by adopting a similar methodology to that of \cite{Multi_nominal}, which extends this approach using multinomial probit regression for BART.

Another potential extension is in survival analysis, where a probit link function could be employed following the approach outlined in \cite{Survival_Analysis}, which adapts BART to model time-to-event data. In that framework, survival times are transformed into a latent probit regression setup, where an underlying continuous variable determines whether an event has occurred by a given time. This allows the model to account for censoring while maintaining flexibility in capturing nonlinear effects. 

\begin{center}
{\large\bf SUPPLEMENTARY MATERIAL}
\end{center}

\begin{description}

\item[GitHub Respiratory:] a repository on GitHub with all the code and data sets to run the algorithm and produce all the figures in this paper.\bf{ \href{https://github.com/Adam-Stone2/Binary_AddiVortes}{AddiVortes R code and data sets}.}

\end{description}

\bibliography{sn-bibliography}

\end{document}